\documentclass [11pt]{article}
\usepackage{amsmath}
\usepackage{graphicx}
\usepackage{amsfonts}
\usepackage{amssymb}
\usepackage{stmaryrd}
\usepackage{amscd}
\newcommand{\bc}{\begin{center}}
\newcommand{\ec}{\end{center}}
\newcommand{\be}{\begin{equation}}
\newcommand{\ee}{\end{equation}}
\newcommand{\bea}{\begin{eqnarray}}
\newcommand{\eea}{\end{eqnarray}}
\newcommand{\ba}{\begin{array}}
\newcommand{\ea}{\end{array}}
\newcommand{\edc}{\end{document}}

\def\f{\varphi}

\def\O{\Omega}

\def\s{\sigma}

\def\C{{\cal C}}

\begin{document}
\thispagestyle{empty}
\begin{center}

{\bf GROUND STATES AND GIBBS MEASURES OF ISING MODEL WITH COMPETING INTERACTIONS AND AN EXTERNAL FIELD ON A CAYLEY TREE}\\
\vspace{0.4cm}
M.M. Rahmatullaev \footnote{mrahmatullaev@rambler.ru}\\
{\it V.I. Romanovskiy Institute of Mathematics,
Uzbekistan Academy of Sciences, 9, University str.,
100174, Tashkent, Uzbekistan; Namangan state university, 316, Uychi str., Namangan,
 Uzbekistan}\\
 M.A. Rasulova \footnote{m\_rasulova\_a@rambler.ru}\\
{\it V.I. Romanovskiy Institute of Mathematics,
Uzbekistan Academy of Sciences, 9, University str.,
100174, Tashkent, Uzbekistan; Namangan state university, 316, Uychi str., Namangan,
 Uzbekistan}\\
J.N. Asqarov \footnote{askarovjavokhir0430@gmail.com\\ The work was supported by the fundamental project (number: F-FA-2021-425) of
The Ministry of Innovative Development of the Republic of Uzbekistan.}\\
{\it Namangan state university, 316, Uychi str., Namangan,
 Uzbekistan}\\

\vspace{0.5cm}

{\bf Abstract}

\end{center}

We consider the Ising model with competing interactions and a nonzero external field on the Cayley tree of order two. We describe ground states and verify
the Peierls condition for the model. Using a contour argument we show the existence of two different Gibbs measures.


\section{INTRODUCTION}

\large  One of the key problems
related to the spin models is the description of the set of Gibbs
measures (see [1], [3], [8], [11], [13], [18]). This problem has a good connection with the problem of
the description the set of ground states. Since the phase
diagram of Gibbs measures (see [8], [13] for details) is close to
the phase diagram of the ground states for sufficiently small
temperatures.

The Ising model, with two values of spin $\pm 1$  was considered
in [11] (see, for example,  [1]-[7], [9], [10], [12], [14], [16], [17]).

In [15], [2], [19] for the classical models (Ising, Potts, SOS) of the statistical mechanics the contour argument (also known as Pirogov-Sinai theory) on Cayley trees was developed.

In this paper, we consider all ground states for
the Ising model with competing interactions and an external field on the Cayley tree of order two and verify the Peierls condition for the model. Using the ground states we also define a notion of contours which allows us to develop a contour argument on the Cayley tree.

The paper is organized as follows. In Section 1 we give some definitions and known facts. Section 2 we study of all (periodic and non-periodic) ground states of the model. In Section 3 we study periodic ground states of the Ising model with competing interactions and a periodic external field. In Section 4 we verify the Peierls condition for the Ising model with competing interactions and an external field. In Section 5 by a contour argument we show the existence of two different Gibbs measures for the model on the Cayley tree of order two.

\textbf{The Cayley tree}. The Cayley tree $\Gamma^k$ (See [1], [5]) of
order $ k\geq 1 $ is an infinite tree, i.e., a graph without
cycles, from each vertex of which exactly $ k+1 $ edges issue. Let
$\Gamma^k=(V, L, i)$, where $V$ is the set of vertices of $
\Gamma^k$, $L$ is the set of edges of $ \Gamma^k$ and $i$ is the
incidence function associating each edge $l\in L$ with its
endpoints $x,y\in V$. If $i(l)=\{x,y\}$, then $x$ and $y$ are
called {\it nearest neighboring vertices}, and we write $l=\langle x,y\rangle$.

 The distance $d(x,y), x,y\in V$ on the Cayley tree is defined by the formula
$$
d(x,y)=\min\{ d | \exists x=x_0,x_1,...,x_{d-1},x_d=y\in V \ \
\mbox{such that}  \ \
 \langle x_0,x_1\rangle,...,\langle x_{d-1},x_d\rangle \}.$$

For the fixed $x^0\in V$ we set $$W_n=\{x\in V \mid  \  d(x,x^0)=n\}, V_n=\{x\in V\mid \ d(x, x^0)\leq n\},$$
$$L_n=\{l=\langle x, y\rangle \in L \mid x, y\in V_n\}.$$
Denote $|x|=d(x,x^0)$, $x\in V$.

A collection of the pairs $\langle x,x_1\rangle,...,\langle x_{d-1},y \rangle$ is called a
{\sl path} from $x$ to  $y$ and we write $\pi(x,y)$ .
 We write $x\prec y$ if
the path from $x^0$ to $y$ goes through $x$.

It is known (see [5]) that there exists a one-to-one
correspondence between the set  $V$ of vertices  of the Cayley
tree of order $k\geq 1$ and the group $G_{k}$ of the free products
of $k+1$ cyclic  groups $\{e, a_i\}$, $i=1,...,k+1$ of the second
order (i.e. $a^2_i=e$, $a^{-1}_i=a_i$) with generators $a_1,
a_2,..., a_{k+1}$.

Denote $S(x)$ the set of direct successors of $x\in G_k$. Let
$S_1(x)$ be denotes the set of all nearest neighboring vertices of
$x\in G_k,$ i.e. $S_1(x)=\{y\in G_k: \langle x,y\rangle\}$ and $ \{x_{\downarrow}\}
=S_1(x)\setminus S(x)$.

\textbf{ The model.} We shall give main definitions and facts
about the model which we are going to study (see [1] for details).
Consider models where the spin takes values in the set
$\Phi=\{-1,1\}$. For $A\subseteq V$ a spin {\it configuration}
$\s_A$ on $A$ is defined as a function
 $x\in A\to\s_A(x)\in\Phi$; the set of all configurations coincides with
$\Omega_A=\Phi^{A}$. Denote $\O=\O_V$ and $\s=\s_V.$ Also put
$-\s_A=\{-\s_A(x), x\in A\}.$  Define a {\it periodic
configuration} as a configuration $\s\in \O$ which is invariant
under a subgroup of shifts $G^*_k\subset G_k$ of finite index.

\textbf{ Definition 1.} A configuration $\s \in \Omega $ is called $G^*_k$
-periodic, if $\s (yx)=\s (x)$ for any $x\in G_k$ and $y\in G^*_k$.

For a given periodic configuration  the index of the subgroup is
called the {\it period of the configuration}.

\textbf{ Definition 2.} A configuration that is invariant with respect to all
shifts is called {\it translation-invariant}.

Let $G_k/G_k^*=\{H_1,...,H_r\}$ be factor group, where $G_k^*$ is a
normal subgroup of index $r\geq 1$.

\textbf{ Definition 3.} Configuration $\sigma(x), x\in
V$ is called  $G_k^*$ -{\it weakly periodic}, if $\sigma(x)=\sigma_{ij}$ for  $x_\downarrow\in H_i, x\in H_j, \forall x\in G_k$.

The Hamiltonian  of the Ising model with competing interactions and an external field
has the form
$$
H(\sigma)=J_1\sum_{\langle x,y\rangle\in
L}\sigma(x)\sigma(y)+J_2\sum_{ x,y\in
V:\atop d(x,y)=2
}\sigma(x)\sigma(y)+\alpha\sum_{x\in V}\sigma(x),  \eqno (1)
$$
where $J_1, J_2,\alpha\in \mathbb{R}$, $\alpha\neq0$ and $\s\in \Omega$.

For a pair of configurations $\sigma$ and $\varphi$ that coincide almost everywhere, i.e. everywhere except for a finite number of
positions, we consider a relative Hamiltonian $H(\sigma,\varphi)$, the difference between the energies of the configurations $\sigma$ and $\varphi$ has the form
$$
H(\sigma,\varphi)=J_1\sum_{\langle x,y\rangle\in L}(\sigma(x)\sigma(y)-\varphi(x)\varphi(y))+ J_2\sum_{x,y \in V :\atop d(x,y)=2}(\sigma(x)\sigma(y)-\varphi(x)\varphi(y))+\alpha\sum_{x\in V}(\sigma(x)-\varphi(x)),\eqno (2)
$$
where $J=(J_1,J_2,\alpha)\in \mathbb{R}^3$ is an arbitrary fixed parameter.

\section{GROUND STATES}

Let $M$ be the set of unit balls  with vertices in $V$. We call
the restriction of a configuration $\s$ to the ball $b\in M$ a
{\it bounded configuration} $\s_b$. We let $c_b$ denote the center of the unit ball $b$.

Define the energy of a configuration $\s_b$ on $b$  by
$$
U(\sigma_b)=\frac{1}{2} J_1\sum_{x\in
S_1(c_b)}\sigma(x)\sigma(c_b)+J_2\sum_{ x,y\in
b:\atop d(x,y)=2}\sigma(x)\sigma(y)+\alpha \sigma(c_b),  \eqno (3)
$$
where $J=(J_1, J_2 ,\alpha)\in \mathbb{R}^3.$

The Hamiltonian (1) can be written as
$$H(\sigma)=\sum_{b\in M}U(\sigma_b).\eqno (4)$$

For any set $A$ we denote by $|A|$ the number of elements in $A$.

{\bf Lemma 1.} For any configuration $\s_b$ we have
$$U(\s_b)\in \{ U_{-,k+1},...,U_{-,0},U_{+,0},..., U_{+,k+1}\},$$
 where
$$U_{\pm,  i}=\bigg(\frac{k+1}{2}-i\bigg)J_1+\bigg(\frac{k(k+1)}{2}+2i(i-k-1)\bigg)J_2\pm \alpha,
\ \ i=0,1,...,k+1.$$
 Let $\C_{\pm, i}=\big\{\f_b: U(\f_b)=U_{\pm, i},  i=0,...,k+1$\}.

{\bf Definition 4.} A  configuration $\f(x),x \in V$ is called a {\it ground
state} for the Hamiltonian (1), if
$$ U(\f_b)=\min\{ U_{-,k+1},...,U_{-,0},U_{+,0},..., U_{+,k+1}\},$$
for any $b\in M$.
\vskip 0.2 truecm
The quantity $U_i(J_1,J_2,\alpha)$ is a linear function of the parameter $({J_1,J_2,\alpha})\in
\mathbb{R}^3. $ For every fixed $m=0,1,...,k+1$ we denote
$$A_{\pm,  m}=\{(J_1,J_2,\alpha)\in \mathbb{R}^3: U_{\pm, m}(J_1,J_2,\alpha)=\min\{U_{-,k+1},...,U_{-,0},U_{+,0},..., U_{+,k+1}\}\}.  \eqno (6)$$
It is easy to check that
$$A_{\pm, 0}=\{(J_1,J_2,\alpha)\in \mathbb{R}^3: J_1\leq 0,\ \  J_1+2kJ_2\leq 0,\ \ \pm \alpha\leq 0\};$$
$$A_{\pm, m}=\{(J_1,J_2,\alpha)\in \mathbb{R}^3: J_2\geq 0,\ \  2(2m-k-2)J_2\leq J_1
\leq 2(2m-k)J_2,\ \  \pm \alpha\leq 0\}, \ \ m=1,2,...,k; $$
$$A_{\pm,k+1}=\{(J_1,J_2,\alpha)\in \mathbb{R}^3: J_1\geq 0,\ \  J_1-2kJ_2\geq 0,\ \  \pm \alpha\leq 0\}$$
and $\mathbb{R}^3=\cup_{i=0}^{k+1}(A_{+,i}\cup A_{-,i}) .$

{\bf The case $k=2$.} It is easy to see that, for any configuration $\s_b$ we have

$$U(\s_b)\in \{{U_{+,0},U_{-,0},U_{+,1},U_{-,1},U_{+,2},U_{-,2},U_{+,3},U_{-,3}}\}$$

where
$$U_{+,0}=\frac {3J_1}{2}+{3J_2}+\alpha,\ \ U_{-,0}=\frac {3J_1}{2}+{3J_2}-\alpha,\ \ U_{+,1}=\frac {J_1}{2}-{J_2}+\alpha,$$
$$U_{-,1}=\frac {J_1}{2}-{J_2}-\alpha, \ \ U_{+,2}=\frac {-J_1}{2}-{J_2}+\alpha, \ \ U_{-,2}=\frac {-J_1}{2}-{J_2}-\alpha,$$
$$U_{+,3}=\frac {-3J_1}{2}+{3J_2}+\alpha, \ \ U_{-,3}=\frac {-3J_1}{2}+{3J_2}-\alpha.$$
Using (6), we obtain
$$A_{+,0}=\{(J_1,J_2,\alpha)\in \mathbb{R}^3: J_1\leq 0,\ \  J_1+4J_2\leq 0,\ \ \alpha\leq 0\}, $$
$$A_{-,0}=\{(J_1,J_2,\alpha)\in\mathbb{R}^3: J_1\leq 0,\ \  J_1+4J_2\leq 0,\ \ \alpha\geq 0\}, $$
$$A_{+,1}=\{(J_1,J_2,\alpha)\in \mathbb{R}^3: J_1\leq 0,\ \  J_1+4J_2\geq 0,\ \ \alpha\leq 0\}, $$
$$A_{-,1}=\{(J_1,J_2,\alpha)\in \mathbb{R}^3: J_1\leq 0,\ \  J_1+4J_2\geq 0,\ \ \alpha\geq 0\}, $$
$$A_{+,2}=\{(J_1,J_2,\alpha)\in \mathbb{R}^3: J_1\geq 0,\ \  J_1-4J_2\leq 0,\ \ \alpha\leq 0\}, $$
$$A_{-,2}=\{(J_1,J_2,\alpha)\in \mathbb{R}^3: J_1\geq 0,\ \  J_1-4J_2\leq 0,\ \ \alpha\geq 0\}, $$
$$A_{+,3}=\{(J_1,J_2,\alpha)\in \mathbb{R}^3: J_1\geq 0,\ \  J_1-4J_2\geq 0,\ \ \alpha\leq 0\}, $$
$$A_{-,3}=\{(J_1,J_2,\alpha)\in \mathbb{R}^3: J_1\geq 0,\ \  J_1-4J_2\geq 0,\ \ \alpha\geq 0\}$$
and $\mathbb{R}^3=\cup_{i=0}^{k+1}(A_{+,i}\cup A_{-,i}).$

Let $A\subset \{1,2,...,k+1\}$, $H_A=\{x\in G_k: \sum_{j\in
A}w_j(x)-$even$\},$ where $w_j(x)$ is the number of letters $a_j$
in the word $x.$ Note that $H_A$ is a normal subgroup of index two (see [1]).
Let $G_k/H_A=\{H_A,G_k\setminus H_A\}$ be the quotient group. We set $H_0:=H_0(A)=H_A, H_1:=H_1(G_k\setminus H_A)=G_k\setminus H_A$.

The $H_A$-periodic configurations are of the form  $$
\varphi(x)= \left\{%
\begin{array}{ll}
    l_{0}, & \textrm{if} \ x \in H_0, \\
    l_{1}, & \textrm{if} \ x \in H_1, \\
\end{array}%
\right. \eqno(7)$$ \qquad   \\
where $l_{i} \in \Phi, i \in \{0,1\}.$

 \vskip 0.2 truecm

 {\bf Theorem 1.} \textit{ Let $k=2$. The following assertions hold for the Ising model with competing interactions and an external field $\alpha\neq0$:}

 1.\textit { (i) If $(J_1,J_2,\alpha)\in A_{+,0}$, then $\f(x)=1$, $\forall x \in V$ configuration is translation-invariant ground state.}

 \textit {(ii) If $(J_1,J_2,\alpha)\in A_{-,0}$, then $\f(x)=-1$, $\forall x \in V$ configuration is translation-invariant ground state.}

 2.\textit{ All $H_A$-periodic ground states are translation-invariant.}

{\bf Proof.} 1. {(i) If \ $\sigma(x)=1$ \ \ $\forall x \in V$, then we have $\sigma_b \in C_{+,0}$ $\forall b \in M$. Consequently $\forall x \in V$, $\sigma(x)=1$ configuration is translation-invariant ground state on the set $A_{+,0}$.

(ii) If \ $\sigma(x)=-1$ \ \  $\forall x \in V$, then we have $\sigma_b \in C_{-,0}$  $\forall b \in M$. Consequently $\forall x \in V$, $\sigma(x)=-1$ configuration is translation-invariant ground state on the set $A_{-,0}$.\\
2. Let $|A|=1$ and

$$\varphi(x)=\left\{%
\begin{array}{ll}
    +1, \ \ \textrm{if} \ x \in H_0, \\
    -1, \ \ \textrm{if} \ x \in H_1. \\
\end{array}%
\right. $$ \qquad
Let $B_\pm=\{x\in S_1(c_b): \f_b(x)=\pm1\}$. \\
If $c_b \in H_0$, then we have $$\f(c_b)=1,\ \ |B_+|=2, \ \ |B_{-}|=1 ,$$ thus $\f_b \in C_{+,1}$.\\
If $c_b \in H_1$, then we have $$\f(c_b)=-1,\ \ |B_{-}|=2, \ \ |B_+|=1 ,$$ thus $\f_b \in C_{-,1}$.\\
Consequently, $\f$ configuration is ground state on the set
$$A_{+,1}\cap A_{-,1}=\{(J_1,J_2,\alpha)\in \mathbb{R}^3:J_1\leq 0,\ \  J_1+4J_2\leq 0,\ \ \alpha=0\}.$$
However, external field should be  only $\alpha \neq0$ due to the assumption of the theorem. The remaining cases (in particular, the case $|A|=2$) can be checked similarly. The theorem is proved.

Recall that model (1) coincides with the Ising model with competing interactions when $\alpha=0$. Periodic ground states for this model were studied in [2], weakly periodic ground states were studied in [16], [17].

The $H_A$-weakly periodic configurations are of the form  $$
\varphi(x)= \left\{%
\begin{array}{ll}
    l_{00}, & \textrm{if} \ \ {x_{\downarrow} \in H_0}, \ x \in H_0, \\
    l_{01}, & \textrm{if} \ \ {x_{\downarrow} \in H_0}, \ x \in H_1, \\
    l_{10}, & \textrm{if} \ \ {x_{\downarrow} \in H_1}, \ x \in H_0, \\
    l_{11}, & \textrm{if} \ \ {x_{\downarrow} \in H_1}, \ x \in H_1, \\
\end{array}%
\right. \eqno(8)$$ \qquad   \\
where $l_{ij} \in \Phi, i,j \in \{0,1\}.$

{\bf Theorem 2.} \textit {Let $k=2$. All $H_A$-weakly ground states are translation-invariant for the Ising model with competing interactions and an external field $\alpha\neq0$.}

{\bf Proof.} Let $|A|=1$ and\\
$$
\varphi(x)= \left\{%
\begin{array}{ll}
    -1, & \textrm{if} \ \ {x_{\downarrow} \in H_0}, \ x \in H_0, \\
    +1, & \textrm{if} \ \ {x_{\downarrow} \in H_0}, \ x \in H_1, \\
    +1, & \textrm{if} \ \ {x_{\downarrow} \in H_1}, \ x \in H_0, \\
    -1, & \textrm{if} \ \ {x_{\downarrow} \in H_1}, \ x \in H_1. \\
\end{array}%
\right. $$ \qquad   \\
Assume $c_b \in H_1$. The possible cases are:\\
a) $c_{b_{\downarrow}}\in H_0$ and $\varphi_b(c_{b_{\downarrow}})=-1$, then
$$\varphi_b(c_{b})=1,\ \ |B_+|=0,\ \ |B_{-}|=3 \ \ \textrm{and} \ \ \varphi_b\in C_{+,3}.$$
b) $c_{b_{\downarrow}}\in H_1$ and $\varphi_b(c_{b_{\downarrow}})=-1$, then
$$\varphi_b(c_{b})=-1,\ \ |B_+|=1,\ \ |B_{-}|=2 \ \ \textrm{and} \ \ \varphi_b\in C_{-,1}.$$
c) $c_{b_{\downarrow}}\in H_1$ and $\varphi_b(c_{b_{\downarrow}})=+1$, then
$$\varphi_b(c_{b})=-1,\ \ |B_+|=2,\ \ |B_{-}|=1 \ \ \textrm{and} \ \ \varphi_b\in C_{-,2}.$$
Therefore $A_{+,3}\cap A_{-,1}\cap A_{-,2}=\{(0,0,0)\}.$ However, on the obtained set $\varphi$ is not ground state since its external field is $\alpha \neq0$ in the assumption of the theorem. The remaining cases can be checked similarly. The theorem is proved.

{\bf Theorem 3.} \textit {For the Ising model with competing interaction and external field $\alpha\neq0$ configuration $\varphi$ is ground state if and only if  it is translation-invariant gound state.}

{\bf Proof.} If we consider non-translation-invariant configuration $\sigma$ on the Cayley tree, for $b_1,b_2\in M$, we have $U{(\sigma_
{b_1})}=U_{+,i}$ or $U{(\sigma_{b_2})}=U_{-,j}$ , $ \ i,j \in \{0,1,2,3\}.$ Consequently, this configuration is ground state on the set $A_{+,i}\cap A_{-,j}\subset\{(J_1,J_2,\alpha)\in \mathbb{R}^3, \alpha=0\}.$ This contradicts the condition $\alpha \neq0$. The theorem is proved.

{\bf Remark 1.} Note that in [2] periodic ground states for the Ising model with two step interactions and zero external field on the Cayley tree were described. In [16], [17] weakly periodic ground states for the Ising model with competing interactions and zero external field were described. In [14] all ground states for the Ising model with non-zero external field were described.

The Theorem 3 asserts that if the nonzero external field is constant, then all ground states for the Ising
model with competing interactions are translation-invariant. Now we shall study
$G_k^{(2)}$-periodic ground states for the Ising model with competing interactions and
$G_k^{(2)}$-periodic external field.

\section{MODEL WITH A PERIODIC EXTERNAL FIELD}

The Ising model with the competing interactions and $G_k^{(2)}$-periodic external field is defined by the Hamiltonian:

$$
H(\sigma)=J_1\sum_{\langle x,y\rangle\in
L}\sigma(x)\sigma(y)+J_2\sum_{ x,y\in
V:\atop d(x,y)=2
}\sigma(x)\sigma(y)+\sum_{x\in V}\alpha_x\sigma(x),\eqno(9)
$$
where $J_1,J_2, \alpha_{x} \in \mathbb{R}$ and
$$
 \alpha_x=\left\{ \begin{array}{ll}
 \alpha_0, \, \mbox{if} \ \ x\in G_k^{(2)},\\[2mm]
 \alpha_1,\,  \mbox{if} \ \ x\in G_k\setminus G_k^{(2)},  \\
 \end{array} \right. \eqno(10)
$$
where $\alpha_0\neq \alpha_1$ and $G_k^{(2)}=\{x\in G_k: |x| \ \
\mbox{is even}\}$.

The energy of a configuration $\sigma_b$ on $b$ is defined by the formula
$$
U(\sigma_b)=\frac{1}{2} J_1\sum_{x\in
S_1(c_b)}\sigma(x)\sigma(c_b)+J_2\sum_{ x,y\in
b:\atop d(x,y)=2}\sigma(x)\sigma(y)+\alpha_{c_b} \sigma(c_b).\eqno (11)
$$
It is not difficult to prove the following lemma.

{\bf Lemma 2.} We have $$U(\sigma_b)\in \{U_{+0}^{(0)},
U_{-,0}^{(0)}, U_{+,0}^{(1)}, U_{-,0}^{(1)}, ..., U_{+,k+1}^{(0)},
U_{-,k+1}^{(0)}, U_{+,k+1}^{(1)}, U_{-,k+1}^{(1)}\}$$ for all
$\sigma_b$, where
$$U_{\pm,
i}^{(j)}=(\frac{k+1}{2}-i)J_1+(\frac{k(k+1)}{2}+2i(i-k-1))J_2\pm\alpha_j, \eqno (12)$$
$i=0, 1, ..., k+1, j=0,1.$

{\bf Definition 5.} A configuration $\varphi$ is called
a ground state of the Hamiltonian (9), if
$$U(\varphi_b)=\min \{U_{+,0}^{(0)},
U_{-,0}^{(0)}, U_{+,0}^{(1)}, U_{-,0}^{(1)}, ..., U_{+,k+1}^{(0)},
U_{-,k+1}^{(0)}, U_{+,k+1}^{(1)}, U_{-,k+1}^{(1)}\}\eqno (13)$$ for all
$b\in M$.

For a fixed $m=0, 1, 2,..., k+1, j=0, 1$, we
set $$ A_{\pm,m}^{(j)}=\{(J_1,J_2, \alpha_0, \alpha_1)\in \mathbb{R}^4 :
U_{\pm,m}^{(j)}=\min \{U_{+,0}^{(0)}, U_{-,0}^{(0)},
...,U_{+,k+1}^{(1)}, U_{-,k+1}^{(1)}\}\}.\eqno (14)$$

Let $k=2$. Quite cumbersome, but not difficult calculations show that
$$
A_{\pm, 0}^{(0)}=\{(J_1,J_2, \alpha_0, \alpha_1) \in \mathbb{R}^4 : \
J_1\leq0,  \\J_1+4J_2\leq0, \mp\alpha_0\geq0,  |\alpha_1|\leq\mp\alpha_0\},
$$
$$
A_{\pm, 0}^{(1)}=\{(J_1,J_2, \alpha_0, \alpha_1) \in \mathbb{R}^4 : \
J_1\leq0, \\J_1+4J_2\leq0, \mp\alpha_1\geq0, |\alpha_0|\leq\mp\alpha_1\},
$$
$$
A_{\pm, 1}^{(0)}=\{(J_1,J_2, \alpha_0, \alpha_1) \in \mathbb{R}^4 : \
J_1\leq0, \\J_1+4J_2\geq0, \mp\alpha_0\geq0, |\alpha_1|\leq\mp\alpha_0\},
$$
$$
A_{\pm, 1}^{(1)}=\{(J_1,J_2, \alpha_0, \alpha_1) \in \mathbb{R}^4 : \
J_1\leq0, \\J_1+4J_2\geq0, \mp\alpha_1\geq0, |\alpha_0|\leq\mp\alpha_1\},
$$
$$
A_{\pm, 2}^{(0)}=\{(J_1,J_2, \alpha_0, \alpha_1) \in \mathbb{R}^4 : \
J_1\geq0, \\J_1-4J_2\leq0, \mp\alpha_0\geq0, |\alpha_1|\leq\mp\alpha_0\},
$$
$$
A_{\pm, 2}^{(1)}=\{(J_1,J_2, \alpha_0, \alpha_1) \in \mathbb{R}^4 : \
J_1\geq0, \\J_1-4J_2\leq0, \mp\alpha_1\geq0, |\alpha_0|\leq\mp\alpha_1\},
$$
$$
A_{\pm, 3}^{(0)}=\{(J_1,J_2, \alpha_0, \alpha_1) \in \mathbb{R}^4 : \
J_1\geq0, \\J_1-4J_2\geq0, \mp\alpha_0\geq0, |\alpha_1|\leq\mp\alpha_0\},
$$
$$
A_{\pm, 3}^{(1)}=\{(J_1,J_2, \alpha_0, \alpha_1) \in \mathbb{R}^4 : \
J_1\geq0, \\J_1-4J_2\geq0, \mp\alpha_1\geq0, |\alpha_0|\leq\mp\alpha_1\}
$$
and $$\mathbb{R}^4=\bigcup_{m=0}^{3}( A_{ \pm, m}^{(0)}\cup A_{ \pm, m}^{(1)}).$$

\textbf{Theorem 4.}
\textbf{a)} \textit{if $(J_1,J_2,\alpha_0,\alpha_1)\in A_{+,3}^{(0)}\cap
A_{-,3}^{(1)}$, then $G_2^{(2)}$-periodic configuration
$$
 \sigma(x)=\left\{ \begin{array}{ll}
 1, \,  \ \ \textrm{if} \,   \ \ x \in G_k^{(2)},\\[2mm]
 -1,\, \textrm{if} \, \ \ x \in G_k\setminus G_k^{(2)}\\
\end{array} \right.\eqno (15)
$$
is a $G_2^{(2)}$-periodic ground
state for the model (9);}

\textbf{b)} \textit{if $(J_1,J_2,\alpha_0,\alpha_1)\in A_{-,3}^{(0)}\cap
A_{+,3}^{(1)}$, then $G_2^{(2)}$-periodic configuration
$$
 \sigma(x)=\left\{ \begin{array}{ll}
 -1, \, \textrm{if} \, \ \ x \in G_k^{(2)},\\[2mm]
 1,\,  \ \ \textrm{if} \, \ \ x \in G_k\setminus G_k^{(2)}\\
\end{array} \right.\eqno (16)
$$
is a $G_2^{(2)}$-periodic ground
state for the model (9).}\\

{\bf Proof. a)} We consider (15). Then we have $\sigma (c_b)=1$ or $\sigma (c_b)=-1$, $\forall b \in M$. If $\sigma (c_b)=1$ then $\forall x \in S_1(c_b)$
we have $\sigma (x)=-1$. In this case by Lemma 2 we take $$U_{+,3}^{(0)}=\frac {-3J_1}{2}+{3J_2}+\alpha_0.$$
If $\sigma (c_b)=-1$ then $\forall x \in S_1(c_b)$
we have $\sigma (x)=+1$. In this case by Lemma 2 we take $$U_{-,3}^{(1)}=\frac {-3J_1}{2}+{3J_2}-\alpha_1.$$
From these cases, it follows that the $G_2^{(2)}$-periodic configuration $\sigma(x), x\in V$ defined in (15) is a ground state on the set
$$
A_{+,3}^{(0)}\cap
A_{-,3}^{(1)}=\{(J_1,J_2, \alpha_0, \alpha_1) \in \mathbb{R}^4 : \
J_1\geq0, \\J_1-4J_2\geq0, \alpha_0\leq0, \alpha_1=-\alpha_0\}.$$

\textbf{b)} We consider (16). If $\sigma (c_b)=+1$ then $\forall x \in S_1(c_b)$
we have $\sigma (x)=-1$. In this case by Lemma 2 we take $$U_{-,3}^{(0)}=\frac {-3J_1}{2}+{3J_2}-\alpha_0.$$
If $\sigma (c_b)=+1$ then $\forall x \in S_1(c_b)$
we have $\sigma (x)=-1$. In this case by (11) we take $$U_{+,3}^{(1)}=\frac {-3J_1}{2}+{3J_2}+\alpha_1.$$
From these cases, it follows that the $G_2^{(2)}$-periodic configuration $\sigma(x), x\in V$ defined in (16) is a ground state on the set
$$
A_{-,3}^{(0)}\cap
A_{+,3}^{(1)}=\{(J_1,J_2, \alpha_0, \alpha_1) \in \mathbb{R}^4 : \
J_1\geq0, \\J_1-4J_2\geq0, \alpha_0\geq0, \alpha_1=-\alpha_{0}\}.$$
The theorem is proved.

\textbf{Theorem 5.} \textit{For the model (9) there is no translation-invariant ground states}.\\
\textbf{Proof. a)} Let $\sigma (x)=1$ $\forall x \in V$. In this case $\forall b \in M$ we have $c_b \in C_{+,0}^{(0)} $ or $c_b \in C_{+,0}^{(1)} $. Thus we obtain that $U(\sigma _b)=U_{+,0}^{(0)}$ or $U(\sigma _b)=U_{+,0}^{(1)}$. The configuration $\sigma (x)=1$ $\forall x \in V$ is a ground state on the set $A_{+,0}^{(0)}\cap
A_{+,0}^{(1)}$. But this contradicts the condition $\alpha_0 \neq \alpha_1 $. \\
\textbf{b)} Let $\sigma (x)=-1$ $\forall x \in V$. In this case $\forall b \in M$ we have $c_b \in C_{-,0}^{(0)} $ or $c_b \in C_{-,0}^{(1)}$. Thus we obtain that $U(\sigma _b)=U_{-,0}^{(0)}$ or $U(\sigma _b)=U_{-,0}^{(1)}$. The configuration $\sigma (x)=-1$ $\forall x \in V$ is a ground state on the set $A_{-,0}^{(0)}\cap
A_{-,0}^{(1)}$. But this contradicts the condition $\alpha_0 \neq \alpha_1 $. The theorem is proved.

\section{THE PEIERLS CONDITION}

{\bf Definition 6.} Let $GS(H)$ be the complete set of all ground states of the Hamiltonian (1). A ball $b\in M$ is said to be an $improper$ ball of the  configuration $\sigma$ if $\sigma_b\neq\varphi_b$ for any $\varphi_b\in GS(H)$. The union of the improper balls of a configuration
$\sigma$ is called the \textit{boundary of the configuration} and denoted by $\partial(\sigma)$.

{\bf Definition 7.} The relative Hamiltonian $H$ defined in (2) with the set of ground states $GS(H)$
satisfies the Peierls condition if for any $\varphi\in GS(H)$ and any configuration $\sigma$ coinciding almost everywhere with $\varphi$,
$$H(\sigma,\varphi)\geq\lambda|\partial(\sigma)|,$$
where $\lambda$ is a positive constant which does not depend on $\sigma$ and $|\partial(\sigma)|$ is the
number of unit balls in $\partial(\sigma)$.

Let $\mathbf{P}=int A_{+,0}\cup int A_{-,0}\equiv\{(J_1,J_2,\alpha)\in \mathbb{R}^3: J_1<0,\ \  J_1+4J_2<0,\ \ \alpha<0\}\cup$$$\cup\{(J_1,J_2,\alpha)\in\mathbb{R}^3: J_1<0,\ \  J_1+4J_2<0,\ \ \alpha>0\}.$$ It is easy to see that
$$\mathbf{P}=\{(J_1,J_2,\alpha)\in \mathbb{R}^3: J_1<0,\ \  J_1+4J_2<0,\ \ \alpha\neq0\}.$$

\textbf{Theorem 6.} \textit{If $(J_1,J_2,\alpha)\in \mathbf{P}$ then the Peierls condition is satisfied.}

\textbf{Proof.} Denote $$\mathbf{U}=\{{U_{+,0},U_{-,0},U_{+,1},U_{-,1},U_{+,2},U_{-,2},U_{+,3},U_{-,3}}\},$$
$U^{\textrm{min}}=\textrm{min} \mathbf{U}$ and $$\lambda_0=\textrm{min}\ \{\mathbf{U}\backslash \{U^{\textrm{min}}\}\}- U^{\textrm{min}}.\eqno (17)$$
Note that
$$
 U^{\textrm{min}}=\left\{ \begin{array}{ll}
 U_{+,0}, \, \mbox{if} \ \ (J_1,J_2,\alpha)\in int A_{+,0},\\[2mm]
 U_{-,0},\,  \mbox{if} \ \ (J_1,J_2,\alpha)\in int A_{-,0}.  \\
 \end{array} \right.
$$

$U_{+,0}=U_{-,0}=U_{+,1}=U_{-,1}=U_{+,2}=U_{-,2}=U_{+,3}=U_{-,3}$ iff $(J_1,J_2,\alpha)=(0,0,0)$. Consequently, $\lambda_0>0$ if
$(J_1,J_2,\alpha)\in \mathbf{P}$. \ \

Suppose $\sigma$ coincides almost everywhere with a ground state $\varphi\in GS(H)$ then
we have $U(\sigma_b)-U(\varphi_b)\geq\lambda_0$ for any $b\in \partial(\sigma)$ since $\varphi$ is a ground state. Thus,
$$H(\sigma,\varphi)=\sum_{b\in
M}(U(\sigma_b)-U(\varphi_b))=\sum_{b\in\partial(\sigma)}(U(\sigma_b)-U(\varphi_b))\geq\lambda_0|\partial(\sigma)|.$$
Therefore, the Peierls condition is satisfied for $\lambda=\lambda_0$. The theorem is proved.

\section{CONTOURS AND GIBBS MEASURES}

Let $\Lambda\subset V$ be a finite set,  $\Lambda^{'}=V\backslash\Lambda$ and $\omega_{\Lambda^{'}}=\{\omega (x), x\in\Lambda^{'}\}$, $\sigma_{\Lambda}=\{\sigma(x),x\in \Lambda\}$ be given configurations. For a finite domain $\Lambda$ with the boundary condition $\omega_{\Lambda^{'}}$ given on its complement $\Lambda^{'}$, the conditional Hamiltonian is

$$H(\sigma_\Lambda| \omega_{\Lambda^{'}})=\sum_{b\in M: \atop b\cap\Lambda\neq\emptyset }U(\sigma_b),\eqno(18)$$
where
$$
 \sigma_b(x)=\left\{ \begin{array}{ll}
 \sigma(x), \, \mbox{if} \,   \ \ x \in b\cap\Lambda,\\[2mm]
 \omega(x),\,  \mbox{if} \, \ \ x \in b\cap\Lambda^{'}.\\
\end{array} \right.
$$

Let $\omega_{\Lambda^{'}}^{\varepsilon}\equiv\varepsilon,\varepsilon=+1$ be a constant configuration outside $\Lambda.$ For a given $\varepsilon$ we extend the configuration $\sigma_{\Lambda}$ inside $\Lambda$ to the Cayley tree by the constant configuration and denote this configuration by $\sigma_{\Lambda}^{\varepsilon}$ and $\Omega_{\Lambda}^{\varepsilon}$ the set of all such configurations.

Now we describe a boundary of the configuration $\sigma_{\Lambda}^{\varepsilon}$. For the sake of simplicity we consider only case $(J_1,J_2,\alpha)\in int A_{+,0}.$ In this case by Theorem 1 we have $GS(H)=\{\sigma^{+}\equiv+1\}.$ Fix $+1$-boundary condition. Put $\sigma_{n}=\sigma^{+}_{V_{n}}$ and $\sigma_{n,b}=(\sigma_{n})_{b}.$ By Definition $6$ the boundary of the configuration $\sigma_{n}$ is
$$\partial\equiv\partial(\sigma_{n})=\{b\in M_{n+2}:\sigma_{n,b}\neq\sigma^{+}_{b}\},$$
where $M_{n}=\{b\in M:b\cap V_{n}\neq\varnothing\}.$

The boundary $\partial$ contains of $2k+3$ parts
$$\partial^{+}_{i}=\{b\in M_{n+2}:\sigma_{n,b}\in \Omega_{i}\}, \ i=1,2,...,k+1;$$
$$\partial^{-}_{i}=\{b\in M_{n+2}:\sigma_{n,b}\in \Omega^{-}_{i}\}, \ i=0,1,2,...,k+1,$$
where $$\Omega_{i}=\{\sigma_b:\sigma_b(c_b)=+1, |\{x\in b\setminus \{c_b\}: \sigma_b(x)=-1\}|=i\},$$
$$\Omega^{-}_{i}=\{-\sigma_b=\{-\sigma_b(x), x\in b\}: \sigma_b\in \Omega_{i}\}.$$

Consider $V_{n}$ and for a given configuration $\sigma_{n}$ (with boundary condition +1) denote
$$V^{-}_{n}\equiv V^{-}_{n}(\sigma_{n})=\{t\in V_{n}:\sigma_n(t)=-1\}.$$
Let $G^{n}=(V_{n}^{-},L_{n}^{-})$ be the graph such that
$$L_{n}^{-}=\{l=\langle x,y\rangle\in L:x,y\in V_{n}^{-}\}.$$

It is clear that for a fixed $n$ the graph $G^{n}$ contains a finite $(=m)$ of maximal connected subgraphs $G_{r}^{n}$, i.e
$$G^{n}=\{G_{1}^{n},...,G_{m}^{n}\}, \ \ G_{r}^{n}=(V_{n,r}^{-},L_{n,r}^{-}), \ \  r=1,...,m,$$
here $V_{n,r}^{-}$ is the set of vertices and $L_{n,r}^{-}$ is the set of edges of $G_{r}^{n}.$

Two edges $l_1,l_2\in L$ are called nearest neighboring edges if $|i(l_{1})\cap i(l_{2})|=1,$ and we write $\langle l_1,l_2\rangle_1.$

For a given graph $G$ denote by $V(G)-$ the set of vertices and by $E(G)-$ the set of edges of $G.$
$$D_{\textrm{edge}}(K)=\{l_1\in L \setminus E(K): \exists l_{2}\in E(K)\ \ \textrm{such} \ \ \textrm{that}\ \ \langle l_1,l_2\rangle_1 \}.$$
The (finite) sets $D_{\textrm{edge}}(G_{r}^{n})$ are called $subcontours$ of the boundary $\partial.$ The set $V_{n,r}^{-}, \ \ r=1, ..., m$ is called the $interior,$ Int$D_{\textrm{edge}}(G_{r}^{n}),$ of $D_{\textrm{edge}}(G_{r}^{n}).$ For any two subcontours $T_1,T_2$ the distance $dist(T_1,T_2)$ is defined by
$$\textrm{dist}(T_1,T_2)=\min_{x\in V(T_1), \atop y\in V(T_2)} d(x,y),$$
where $d(x,y)$ is the distance between $x,y\in V$ (see Section 1).

{\bf Definition 8.} The subcontours $T_1,T_2$ are called $adjacent$ if $\textrm{dist}(T_1,T_2)\leq2.$ A set of subcontours
$\mathcal{A}$ is called connected if for any two subcontours $T_1,T_2\in \mathcal{A}$ there is a collection of subcontours $T_1=\widetilde{T}_1, \widetilde{T}_2, ..., \widetilde{T}_l=T_2$ in the set $\mathcal{A}$ such that for each $i=1, ..., l-1$ the subcontours $\widetilde{T}_i$ and $\widetilde{T}_{i+1}$ are adjacent.

{\bf Definition 9.} Any maximal connected set (component) of subcontours is called contour of the set $\partial.$

The set of edges from a contour $\gamma$ is denoted by supp $\gamma$.

For a given contour $\gamma$ we put
$$imp_{i}^{\varepsilon}\gamma=\{b\in\partial_{i}^{\varepsilon}:b\cap\gamma\neq\emptyset\}, \ \varepsilon=-1, \ i=0, 1,...,k+1; \  \varepsilon=1, \ i=1,...,k+1;$$
$$imp\gamma=(\cup_{i=1}^{k+1}imp_{i}^+\gamma)\cup(\cup_{i=0}^{k+1}imp_{i}^-\gamma);$$
$$|\gamma|=|imp\gamma|, \ |\gamma^{\varepsilon}_{i}|=|imp_{i}^{\varepsilon}\gamma|, \ |\gamma_{i}|=|\gamma_{i}^{+}|+|\gamma_{i}^{-}|.$$
It is easy to see that the collection of contours $\tau=\{\gamma_{r}\}$ generated by the boundary $\sigma_{n}$ has  the following properties:

$(i)$ Every contour $\gamma\in\tau$ lies inside of the set $V_{n+1};$

$(ii)$ For every two contours $\gamma_{1},\gamma_{2}\in \tau$ we have dist$(\gamma_{1},\gamma_{2})>2,$ thus their supports supp$\gamma_{1}$ and  supp$\gamma_{2}$ are disjoint.

A collection of contours $\tau=\{\gamma\}$ that has the properties $(i)$-$(ii)$ is called a configuration of contours. As we have seen, the configuration $\sigma_{n}$ of spin generates the configuration of contours $\tau=\tau(\sigma_{n}).$ The converse assertion is also true. Indeed, for a given collection of contours $\{\gamma_{r}\}^{m}_{r=1}$ we put $\sigma _{n}(x)=-1$ for each $x\in \textrm{Int}\gamma_{r},r=1,...,m$ and $\sigma_{n}(x)=+1$ for each $x\in V_{n}\backslash\cup_{r=1}^{m}\textrm{Int}\gamma_{r}.$

Let us define a graph structure on $M$ (i.e. on the set of all unit balls of the Cayley tree) as follows. Two balls $b,b^{'}\in M$ are connected by an edge if they are neighbors, i.e. have a common edge. Denote this graph by $G(M).$ Note that the graph $G(M)$ is a Cayley tree of order $k\geq1.$ Here the vertices of this graph are balls of $M$.

Denote $N_r(x)=|\{\gamma: x\in \gamma,|\gamma|=r\}|$.

{\bf Lemma 3.} [15] If $k=2$ (i.e. the Cayley tree of order two). Then
$$N_r(x)\leq \textrm{Const} \cdot (4e)^{2r}.\eqno(19)$$

Following lemma gives a contour representation of Hamiltonian.

{\bf Lemma 4.} The energy $H_n(\sigma_n)\equiv H_{V_n}(\sigma_n|\omega_{V'_n}=+1)$ (see (18)) has the form
$$H_n(\sigma_n)=\sum_{i=1}^{k+1}(U_{+,i}-U_{+,0})\cdot |\partial_i^+|+\sum_{i=0}^{k+1}(U_{-,i}-U_{+,0})\cdot |\partial_i^-|+|M_{n+2}|\cdot U_{+,0}.\eqno (20)$$

{\bf Proof.} $$H_n(\sigma_n)=\sum_{b\in M_{n+2}}U(\sigma_{n,b})=\sum_{i=1}^{k+1}U_{+,i}\cdot |\partial_{i}^{+}|+\sum_{i=0}^{k+1} U_{-,i}\cdot|\partial_{i}^{-}|+$$ $$+(|M_{n+2}|-\sum_{i=1}^{k+1}|\partial_{i}^{+}|-\sum_{i=0}^{k+1}|\partial_{i}^{-}|)\cdot U_{+,0}.\eqno (21)$$\\
From (21) we get (20). The lemma is proved.

{\bf Lemma 5.} Assume $(J_1,J_2,\alpha)\in IntA_{+,0}.$ Let $\gamma$ be a fixed contour and
$$p_{+}(\gamma)=\frac{\sum_{\sigma_{n}:\gamma\in\partial} \textrm{exp}(-\beta H_{n}(\sigma_{n}))}{\sum_{\tilde{\sigma}_{n}}\textrm{exp}(-\beta H_{n}(\tilde{\sigma}_{n}))} \eqno (22)$$
Then $$p_{+}(\gamma)\leq \textrm{exp}(-\beta\lambda_{0}\mid \gamma\mid),\eqno (23)$$ where $\lambda_{0}$ is defined by formula (17), that is
$$\lambda_{0}=\min\{\mathbf{U}\setminus\{U_{+,0}\}\}-U_{+,0}$$
and $\beta=\frac{1}{T}$, $T>0$ $-$ temperature.

{\bf Proof.} Put $\Omega_{\gamma}=\{\sigma_{n}:\gamma\subset\partial\}, \ \Omega_{\gamma}^{0}=\{\sigma_{n}:\gamma\cap\partial=\varnothing\}$ and define a map $\chi_{\gamma}: \Omega_{\gamma}\rightarrow \Omega_{\gamma}^{0}$ by
$$\chi_{\gamma}(\sigma_{n})(x)=\left\{%
\begin{array}{ll}
    +1, \ \   \ \ \ if & \ x \in \textrm{Int}\gamma, \\
    \sigma_{n}(x), \ \ if & \ x \ \bar{\in} \ \textrm{Int}\gamma. \\
\end{array}%
\right. $$ \qquad
For a given $\gamma$ the map $\chi_{\gamma}$ is a one-to-one map. We need to the following

{\bf Lemma 6.} For any $\sigma_{n}\in \Omega_{V_{n}}$ we have
$$|\partial_{i}^{+}(\sigma_{n})|=|\partial_{i}^{+}(\chi_{\gamma}(\sigma_{n}))|+|\gamma_{i}^{+}|,$$  where $i=\overline{1,k+1},$
$$|\partial_{i}^{-}(\sigma_{n})|=|\partial_{i}^{-}(\chi_{\gamma}(\sigma_{n}))|+|\gamma_{i}^{-}|,$$  where $i=\overline{0,k+1}.$

{\bf Proof.} It is easy to see that the map $\chi_\gamma$ destroys the contour $\gamma$ and all other contours are invariant with respect to $\chi_\gamma$. This completes the proof.  \\
Now we shall continue the proof of Lemma 5. By Lemma 4 we have

$$p_{+}(\gamma)=\frac{\sum_{\sigma_{n}\in\Omega_\gamma} \textrm{exp}(-\beta(\sum_{i=1}^{k+1}(U_{+,i}-U_{+,0})\cdot |\partial_{i}^{+}(\sigma_{n})|+\sum _{i=0}^{k+1}(U_{-,i}-U_{+,0})\cdot|\partial_{i}^{-}(\sigma_n)|))}{\sum_{\tilde{\sigma}_{n}}\textrm{exp}(-\beta(\sum _{i=1}^{k+1}(U_{+,i}-U_{+,0})\cdot|\partial_{i}^{+}(\tilde{\sigma}_{n})|+\sum_{i=1}^{k+1}(U_{-,i}-U_{+,0})\cdot|\partial_{i}^{-}(\tilde{\sigma_{n}})|))}$$
$$\leq\frac{\sum_{\sigma_{n}\in\Omega_{\gamma}}\textrm{exp} (-\beta(\sum_{i=1}^{k+1}(U_{+,i}-U_{+,i})\cdot|\partial_{i}^{+}(\sigma_{n})|+\sum_{i=0}^{k+1}(U_{-,i}-U_{+,0})
\cdot|\partial_{i}^{-}(\sigma_{n})|))}
{\sum_{\tilde{\sigma}_{n}\in \Omega_{\gamma}^{0}}\textrm{exp}{{(-\beta(\sum_{i=1}^{k+1}(U_{+,i}-U_{+,0})\cdot
|\partial_{i}^{+}(\tilde{\sigma}_{n})|+\sum_{i=0}^{k+1}(U_{-,i}-U_{+,0})\cdot|\partial_{i}^{-}(\tilde{\sigma}_{n})|))}}}=$$
$$=\frac{\sum_{\sigma_{n}\in\Omega_{\gamma}}\textrm{exp}(-\beta(\sum_{i=1}^{k+1}(U_{+,i}-U_{+,0})\cdot|\partial_{i}^{+}|+\sum_{i=0}^{k+1}
(U_{-,i}-U_{+,0})\cdot|\partial_{i}^{-}(\sigma_{n})|))}{\sum_{\tilde{\sigma}_{n}\in\Omega_{\gamma}}\textrm{exp}(-\beta(\sum_{i=1}^{k+1}
(U_{+,i}-U_{+,0})\cdot|\partial_{i}^{+}(\chi_{\gamma}{(\tilde{\sigma}_{n})})|+\sum_{i=0}^{k+1}(U_{-,i}-U_{+,0})\cdot|\partial_{i}^{-}
(\chi_{\gamma}(\tilde{\sigma}_{n}))|))}. \eqno(24)
$$
Since $(J_1,J_2,\alpha)\in Int A_{+,0}$ by Theorem 1 we have $GS(H)=\{\sigma^{+}\}$ hence $U_{+,i}-U_{+,0}\geq\lambda$ for any $i=1,...,k+1$ and $U_{-,i}-U_{+,0}\geq\lambda_{0}$ for any $i=0,1,...,k+1$. Thus using this fact and Lemma 6 from (24) we get (23). The lemma is proved.

Using Lemmas 13 and 15 by very similar argument of [15] one can prove

\textbf{Theorem 7.} \textit{I. If $(J_1, J_2, \alpha)\in int A_{+,0}$ then for all sufficiently large $\beta$ there is at least one Gibbs measure for the model (1) on Cayley tree of order two.\\
 II. If $(J_1, J_2, \alpha)\in int A_{-,0}$ then for all sufficiently large $\beta$ there is at least one Gibbs measure for the model (1) on Cayley tree of order two.}

\textbf{Remark 2.} Note that in [2] it was found that Gibbs measures exist for the Ising model with competing interactions and zero external field on the Cayley tree. Theorem 7 shows that Gibbs measures exist when the external field is nonzero.

\textbf{Acknowledgments.}

 We thank professor U.A. Rozikov for his helpful comments.

\textbf{Data Availability Statement.}

Not applicable.

\textbf{Conflicts of Interest.} 

The authors declare that they have no conflict of interest.

{\bf References}

1. U.A. Rozikov, Gibbs measures on Cayley trees, World scientific, (2013).

2. U.A.Rozikov, A Constructive Description of Ground States and
Gibbs Measures for Ising Model With Two-Step Interactions on
Cayley Tree, {\it Jour. Statist. Phys.}, {\bf 122}: 217--235 (2006).

3. R.J. Baxter, Exactly  Solved Models in Statistical
Mechanics, (Academic Press, London/New York, 1982).

4. P.M. Bleher, J. Ruiz, V.A. Zagrebnov. On the purity of the
limiting Gibbs state for the Ising model on the Bethe lattice,
{\it Jour. Statist. Phys.}, {\bf 79}: 473--482 (1995).

5. N.N. Ganikhodzhaev, Group representation and automorphisms of the Cayley tree, {\it Dokl. Akad. Nauk Resp.
Uzbekistan}, {\bf 4}: 3--5 (1994).

6. P.M. Bleher, J. Ruiz, R.H. Schonmann, S. Shlosman and V.A.
Zagrebnov, Rigidity of the critical phases on a Cayley tree, {\it
Moscow Math. Journ.}, {\bf 3}: 345--362 (2001).

7.  N.N. Ganikhodjaev and U.A. Rozikov, A description of periodic
extremal Gibbs measures of some lattice models on the Cayley tree,
{\it Theor. Math. Phys.}, {\bf 111}: 480--486 (1997).

8. R.A. Minlos, Introduction to mathematical statistical
physics (University lecture series,  {\bf 19}, 2000)

9. F.M. Mukhamedov, U.A. Rozikov, On Gibbs measures of models
with competing ternary and binary interactions and corresponding
von Neumann algebras.  {\it Jour. of Stat.Phys.}, {\bf 114}:
825--848 (2004).

10.  Kh.A. Nazarov, U.A. Rozikov, Periodic Gibbs measures for the
Ising model with competing interactions, {\it Theor. Math. Phys.},
{\bf 135}: 881--888 (2003)

11. C. Preston, Gibbs states on countable sets (Cambridge
University Press, London 1974).

12.  U.A. Rozikov, Partition structures of the group
representation of the Cayley tree into cosets by finite-index
normal subgroups and their applications to the description of
periodic Gibbs distributions,  {\it Theor. Math. Phys.}, {\bf 112}:
929--933 (1997).

13.  Ya.G. Sinai, Theory of phase transitions: Rigorous
Results (Pergamon, Oxford, 1982).

14. M.M. Rahmatullaev, M.A. Rasulova, Ground states for the Ising model with an external field on the Cayley tree,  {\it Uzbek Mathematical Journal}, {\bf 3}: 147--155 (2018).

15. U.A. Rozikov, On $q$-component models on Cayley tree: contour method,  {\it Letters in Math. Phys.}, {\bf 71}: 27--38 (2005).

16. M.M. Rahmatullaev, Description of weak periodic ground states of Ising model
with  competing  interactions  on  Cayley  tree,  {\it Applied  mathematics  and
Information science. AMIS USA}, {\bf 4(2)}: 237--241 (2010).

17. U.A. Rozikov, M.M. Rahmatullaev, Weakly Periodic Ground States and Gibbs Measures for the Ising
Model with Competing Interactions on the Cayley Tree, {\it Theor. Math. Phys.}, \textbf{160}, No.3: 1292--1300 (2009).

18. U.A. Rozikov, M.M. Rakhmatullaev, R.M. Khakimov, Periodic Gibbs measures for the Potts model in translation-invariant and periodic external fields
on the Cayley tree, {\it Theor. Math. Phys.}, \textbf{210}(1): 135--153 (2022).

19. G.I. Botirov, U.A. Rozikov, Potts model with  competing  interactions  on  the Cayley  tree: The contour method, {\it Theor. Math. Phys.}, \textbf{153}(1): 1423--1433 (2007).
\end{document}